\documentclass[journal]{IEEEtran}
\usepackage{graphicx}
\usepackage{url}
\usepackage{algorithm,algorithmic}
\usepackage{multirow}
\usepackage{color}
\usepackage{tabularx}

\usepackage{cite}
\ifCLASSINFOpdf
\else
\fi
\usepackage{amsmath}
\usepackage{amsfonts}
\begin{document}
\title{Autoregressive Linguistic Steganography Based on BERT and Consistency Coding}

\author{Xiaoyan Zheng and Hanzhou Wu
\thanks{\emph{Corresponding author: Hanzhou Wu (contact email: h.wu.phd@ieee.org)}}
}


\markboth{}%
{}



\maketitle

\begin{abstract}
Linguistic steganography (LS) conceals the presence of communication by embedding secret information into a text. How to generate a high-quality text carrying secret information is a key problem. With the widespread application of deep learning in natural language processing, recent algorithms use a language model (LM) to generate the steganographic text, which provides a higher payload compared with many previous arts. However, the security still needs to be enhanced. To tackle with this problem, we propose a novel autoregressive LS algorithm based on BERT and consistency coding, which achieves a better trade-off between embedding payload and system security. In the proposed work, based on the introduction of the masked LM, given a text, we use consistency coding to make up for the shortcomings of block coding used in the previous work so that we can encode arbitrary-size candidate token set and take advantages of the probability distribution for information hiding. The masked positions to be embedded are filled with tokens determined by an autoregressive manner to enhance the connection between contexts and therefore maintain the quality of the text. Experimental results have shown that, compared with related works, the proposed work improves the fluency of the steganographic text while guaranteeing security, and also increases the embedding payload to a certain extent.
\end{abstract}


\IEEEpeerreviewmaketitle

\section{Introduction}
\IEEEPARstart{W}{ith} the rapid development of wireless communication and social networking services, a lot of people are instantly sharing daily-life feelings and media data by integrating into social networks through mobile terminal devices. This has brought great convenience to covert communication which aims to reliably convey a secret message to the data receiver without arousing the suspicion of the monitor. Steganography, as a means to covert communication \cite{Simmons:paper}, is an art of embedding secret information in an innocent digital carrier such as image and video by slightly modifying the carrier. Since the resulting modified carrier containing secret information will not introduce noticeable artifacts, by sending the modified carrier via an insecure public channel such as social network, the receiver is able to fully reconstruct the secret information. Steganography secures communication using its concealment, becoming more and more important in information security \cite{TII:Paper1, TII:Paper2}. 

Regardless of the past or the present, text is one of the most important information exchange media in everyday life, and communication and conversation between each other needs to rely on text to convey a large amount of information. At the same time, natural language has high robustness in the process of information transmission, even in the case of interference, text can still keep secret transmission in the public channel without distortion. It can be seen from the above two points that it is reasonable and necessary to study the use of text as the carrier of steganography. However, due to the low redundancy of the text itself, compared with image, audio, video and other digital media, the research on linguistic steganography (LS) is more complex and challenging. 

The basic framework of LS is based on the scene description of the prisoners' problem \cite{Simmons:paper}. In order to ensure that \emph{Alice} and \emph{Bob} communicate secretly without being discovered by the guardian \emph{Wendy}, a steganographic system (stego system) was designed based on natural language processing (NLP), the key problems of which are how to embed secret information in a text (i.e., data embedding for \emph{Alice}) and how to extract secret information from the text containing secret information (i.e., data extraction for \emph{Bob}). The most important purpose for a LS system is to conceal the existence of communication to ensure the \emph{unsuspiciousness} of steganographic texts (stego texts). It means that it should not reveal any evidence of hidden information when the stego texts are being analyzed by humans or computers. In other words, it should be impossible for both the human perceptual system and a linguistic steganalyzer to separate the stego texts from natural ones. Under the condition of good concealment, it is generally quite desirable to embed as much information as possible in a text. To count the amount of the embedded information, ``bits per word (BPW)'' has been widely used in LS methods. Mainstream LS methods focus on improving the imperceptibility (corresponding to concealment) and the embedding payload. However, the relationship between them is mutually competitive and contradictory \cite{Chang:CL, Kang:paper}. Enhanced imperceptibility will reduce the embedding payload, and the increase of the embedding payload will require more modifications to the original text, reducing the imperceptibility (which leads to low security). How to achieve a good balance between them is a core problem for designing LS systems.

LS can generally be divided to two categories: modification based and generation based. Traditional stego systems such as \cite{KeithWinstein:paper, Huo:paper, Chiang:paper} embed secret information in a given text (typically also called \emph{cover}) by modifying the cover, which is referred to as the former category. It is required that the modification should not significantly impair the cover text. Early methods mainly alter the format attributes to embed secret information such as character spacing, font size and text line spacing \cite{SHLow:IEEE1995, Brassil:IEEE1999}. Modifying the file structures has also investigated since it leaves the content of the text unchanged and thus has good concealment, but cannot resist against re-editing attacks \cite{Shahreza:2007, Liu:TIFS2017}. Researchers have also tried to use the transformation of text sentence characteristics \cite{Topkara:ACM2006, Kumar:IEEE2016} or vocabulary \cite{Topkara:ACMMMSec2006, Rizzo:ACM2016} to generate the stego text such as the most common synonym substitution \cite{Liu:IEEE2007, Xiang:CMC2018}. In particular, when to use word (or say token) replacement to embed secret information, one should well design the synonym dictionary as well as the information encoding strategy. Although modification based LS has high semantic concealment, the embedding payload is not high. So, it is often necessary to find a good trade-off between them.

Generation based LS is the process of directly generating a stego text based on secret information and a trained language model (LM). During stego text generation, the trained LM will assign a prediction probability to each candidate word in the word pool such that the present output is the ``most suitable'' word maintaining the text quality well while matching secret information. Compared with traditional modification based LS methods, generation based LS methods can provide a higher embedding payload \cite{luo:ihmmsec2017} but the security cannot be well guaranteed \cite{Fang:paper, Guo:paper, Wu:GNNSteganalysis, Yibiao:paper1, Yibiao:paper2}. Even if the modern LM is combined with arithmetic coding, the anti-statistical analysis ability has not been improved, and there is still a poor anti-staganalysis ability \cite{Ziegler:paper}. So, the security needs to be enhanced.

Recently, Ueoka \emph{et al.} \cite{HonaiUeoka:2021} present a novel masked language model (masked LM) based on BERT (short for Bidirectional Encoder Representations from Transformers) \cite{BERT:paper} for realizing LS, which simplifies the rules of thesaurus construction and provides a new perspective for modification based LS. Though the stego texts generated by this method can carry a relatively high payload, the word prediction in masked positions of this method is parallel. It means that the words corresponding to masked positions have a high degree of independence. It is known that every word in a text is closely related to the fluency of the text. If the words have a high degree of independence, it will be easily recognized by the human perceptual system, inspiring the adversary to develop advanced steganalyzers that reduce the security. Therefore, we urgently need to find a better strategy to generate higher-quality stego texts.

In this paper, we propose a novel autoregressive LS algorithm based on the well-known BERT, which uses an autoregressive strategy to generate the stego texts based on the secret information to be hidden. Our main contribution is that we use an autoregressive strategy for the candidate words obtained from the masked LM, fill the masked positions with replaceable words in turn, and then input them to the masked LM to make predictions. We have also improved the information coding strategy, which imitates the mechanism of word selection by native speakers and improves the readability and the authenticity of sentences. Experimental results have shown that, compared with the non-autoregressive method, the proposed method improves the fluency of the stego text while guaranteeing the security, and increases the embedding payload to a certain extent, which verifies the superiority. 

The rest structure of this paper is organized as follows. We review the related work in Section II. Then, we introduce the proposed work in detail in Section III, followed by convinced experimental results and analysis in Section IV. We conclude this work and provide valuable discussion in Section V.

\section{Related Work}
Recently, Ueoka \emph{et al.} \cite{HonaiUeoka:2021} present a novel masked language model (masked LM) based on BERT for realizing LS, which successfully extends LM to modification based LS. Mathematically, given a text $\textbf{x} = \{x_1, x_2, ..., x_n\}\subset V^n$, where $x_i$ is the $i$-th word sampled from a large vocabulary $V$ and $n$ is the total number of words in the text, the method aims to generate such a stego text $\textbf{y} = \{y_1, y_2, ..., y_n\}\subset V^n$ that a secret message $\textbf{b}\subset \{0,1\}^L$ can be extracted from $\textbf{y}$. The data embedding process can be described as follows. First, a certain number of word positions (also called ``masked positions'') are selected from $\textbf{x}$ according to the secret key $\textbf{k}$. Then, by feeding the words in the non-masked positions to the masked LM, a list of candidate words associated with a prediction probability can be determined for each masked position. Thereafter, based on $\textbf{b}$ and the block coding strategy introduced in \cite{HonaiUeoka:2021}, $\textbf{y}$ can be generated by replacing the words in masked positions in $\textbf{x}$ with the suitable words in the corresponding candidate sets. 

In order to successfully extract $\textbf{b}$ from $\textbf{y}$, the data hider and the data receiver should share the secret key $\textbf{k}$ and the masked LM $\mathcal{M}$. The secret key ensures that the data hider and the data receiver can identify the same masked positions. The masked LM $\mathcal{M}$ ensures that the data hider and the data receiver obtain the same prediction probability of each candidate word for each masked position in order that the data hider knows \emph{how to encode a secret stream to a word} and the data receiver knows \emph{how to decode a word to the corresponding secret stream}.

In summary, this method significantly increases the embedding payload compared with the previous modification based LS methods and provides satisfactory security. It also brings the relationship between modification based LS and generation based LS closer to each other. However, there are two shortcomings in this method which can be easily exploited by the adversary to reveal the existence of hidden information. The first one is that the word prediction in masked positions in the method is parallel. In other words, the word prediction of different masked positions will be intuitively independent of each other, i.e., predicting the word in a specific position will not affect the prediction of another one. However, every word is closely related to the text fluency. If the words have a high degree of independence, it will be easily recognized by the human perceptual system, allowing the adversary to develop new detectors reducing the security. Therefore, we urgently need to design a more efficient strategy for word prediction.

\begin{figure*}[!t]
\centering
\includegraphics[width=\linewidth]{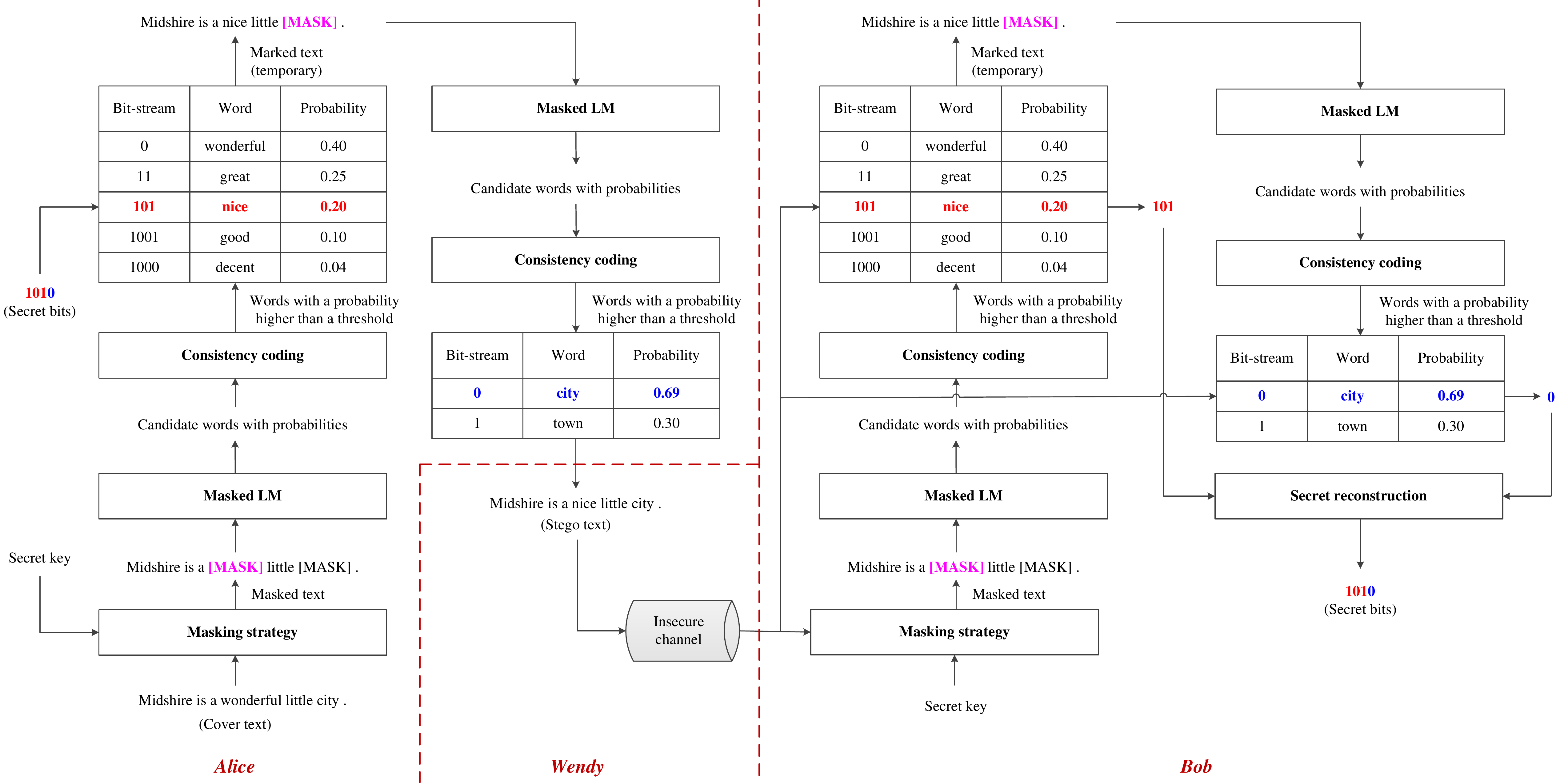}
\caption{An intuitive explanation for the proposed method. \emph{Alice} embeds secret bits in a given cover text by applying the proposed autoregressive LS algorithm based on masked LM (BERT) and consistency coding. \emph{Wendy} detects whether the stego text contains secret bits or not. \emph{Bob} reconstructs the secret bits from the stego text by the operation similar to \emph{Alice}. \emph{Alice} and \emph{Bob} should share some side information including the threshold controlling the number of candidate words, the secret key, masked LM and information encoding strategy in advance, which is required for mainstream LM based LS algorithms as well.}
\label{fig:1}
\end{figure*}

On the other hand, the information encoding strategy (which is a kind of block coding technique) used in this method simply maps a certain number of candidate words to a binary stream with a fixed length for each masked position. For example, for a specific masked position, assuming that there are $m$ words whose prediction probabilities are larger than a threshold $t_p$, the largest $l$ satisfying $2^l\leq m$ can be determined. The above method uses the $2^l$ words with the largest probabilities as the candidate words, each of which carries exactly $l$ secret bits. In this way, during data embedding, the candidate word matching the secret stream with a length of $l$ is used to fill the masked position. It is seen that though the length $l$ can be controlled by tuning the threshold $t_p$, it ignores the probabilistic associations between different words. For example, assuming that $l = 2$ and the candidate words are ``wonderful (0.6)'', ``decent (0.2)'', ``fine (0.1)'' and ``great (0.1)'', where ``wonderful (0.6)'' means that the prediction probability of ``wonderful'' is 0.6, the above method maps each word to a binary stream with a length of 2, e.g., the words can be respectively mapped to ``00'', ``01'', ``10'' and ``11''. Obviously, during data embedding, all the candidate words have the identical probability (i.e., $1/2^l$) to be chosen to fill the masked position. However, according to the masked LM, during data embedding, instead of randomly choosing a word, choosing a word with a prediction probability as high as possible is more desirable for achieving the better quality of the stego text. It means that we urgently need to design a more efficient strategy for information encoding as well.

Word prediction and information encoding have direct impact on the system security. It strongly motivates us to propose a new LS algorithm in this paper, which significantly enhances the security and, more importantly, presents a new perspective to modification based LS. We show the details in Section III. 

\section{Proposed Method}
\subsection{Overview}
As shown in Fig. \ref{fig:1}, the proposed autoregressive LS method involves three participants: the data hider \emph{Alice}, the adversary (attacker) \emph{Wendy}, and the data receiver \emph{Bob}. The goal of \emph{Alice} is to hide a secret message (in the form of a binary stream) into a cover text. The resulting stego text carrying the secret message will be sent to \emph{Bob} via an insecure channel such as the Internet, which is being monitored by the adversary \emph{Wendy} whose purpose is to determine whether a text being transferred contains secret information or not. 

Specifically, according to the secret key and the secret message to be embedded, \emph{Alice} first determines a set of masked positions in the cover text. For each masked position to be processed, a word will be determined according to the masked LM, the present temporary text (which may contain \emph{processed} and \emph{unprocessed} masked positions) and the consistency coding technique so that the word will substitute the masked position and meanwhile match the secret bits to be embedded. When all masked positions are processed, a stego text can be generated and will be sent to \emph{Bob}, who performs the operation similar to \emph{Alice} on the stego text so that the entire secret message can be extracted from the stego text. In order to ensure that the data extraction procedure can be successfully executed, \emph{Alice} and \emph{Bob} should share the necessary side information in advance including the secret key, the threshold controlling the selection of candidate words, the masked LM as well as the consistency coding strategy. In the following, we provide more details.

\subsection{Masking Strategy and Masked Language Model (LM)}
The goal of the masking strategy is to replace some words (tokens) in the cover text with the special token ``[MASK]''. These masked positions will be filled with tokens generated by the masked LM based on the context. The new tokens not only fit into the context, but also carry secret information. It is free for us to design the masking strategy. Though the masking strategy can be carefully crafted, it is not the main contribution of this paper. For fair comparison in experiments, we follow the simple but effective masking strategy in \cite{HonaiUeoka:2021}. In brief summary, a masking interval $f$ (which is an integer and can be considered as a secret key) is used to control the total number of masked positions. A higher $f$ means less tokens are masked, which reduces the embedding capacity but increases the difficulty of detection.

The masked strategy has been previously introduced along with BERT \cite{BERT:paper} as an efficient pretraining strategy for Transformer based nets \cite{transformer:paper}. By using the masked strategy, BERT overcomes the limitations of unidirectional training and realizes the common dependence on context for prediction. Generally speaking, fine-tuning the downstream tasks with the pre-trained BERT according to their specific tasks such as sentence classification, sequence tagging and question answering, can greatly reduce the expense of designing their own architecture. However, fine-tuning is not required in this paper. Different from recurrent neural networks (RNNs) \cite{lstm:paper} processing texts by a word-wise manner and generative pre-trained transformer (GPT) \cite{GPT:paper} obscuring future markers, the masked LM provides superior prediction performance due to its bidirectionality. Therefore, we use the pretrained BERT as the masked LM. 

Let $\mathcal{M}$ and $\mathcal{I} = \{i_1, i_2, ..., i_s\}$ be the masked LM and the index-set for the masked positions determined with the secret key. For each $i\in \mathcal{I}$, the corresponding word $x_i$ in the cover text $\textbf{x} = \{x_1, x_2, ..., x_n\}$ will be replaced with ``[MASK]''. To express it mathematically, for any $\mathcal{S}\subset \mathcal{I}$, we define $\textbf{x}_\mathcal{S}$ as:
\begin{equation}
\textbf{x}_\mathcal{S} = \{x_1', x_2', ..., x_n'\},
\end{equation}
where for $1\leq i\leq n$, we have
\begin{equation}
x_i' = \left\{\begin{matrix}
\text{[MASK]} & \text{if}~i\in\mathcal{S};\\ 
x_i & \text{otherwise}.
\end{matrix}\right.
\end{equation}

Obviously, we have $\textbf{x}_\emptyset = \textbf{x}$ and $\textbf{x}_\mathcal{I}$ is the text after replacing all the words in the masked positions with ``[MASK]''. To embed secret information in the masked positions, we \emph{orderly} process the masked positions. For each masked position, $\mathcal{M}$ will output a prediction probability for each word in the vocabulary according to the temporary text generated previously. A higher probability implies that the corresponding word is more suitable to replace the ``[MASK]'' to fit into the context. Next, we determine the temporary text for each masked position.

Without the loss of generalization, let $i_1 < i_2 < ... < i_s$. The temporary text for the masked index $i_j\in \mathcal{I}$ is determined as follows. First, the temporary text for $i_1 \in \mathcal{I}$, denoted by $\textbf{x}_{[i_1]}$,  is $\textbf{x}_\mathcal{I}$. By feeding $\textbf{x}_{[i_1]}$ to $\mathcal{M}$, we can generate a prediction probability $p_i\in [0, 1]$ for each $v_i\in V = \{v_1, v_2, ..., v_z\}$ and $\sum_{i=1}^{z}p_i = 1$. According to the secret bits to be embedded and the proposed consistency coding technique, we select a word $x_{i_1}^*\in V$ to replace the ``[MASK]'' in the $i_1$-th position in $\textbf{x}$. For $i_2\in I$, the only one difference between $\textbf{x}_{[i_1]}$ and $\textbf{x}_{[i_2]}$ is that in the $i_1$-th position, $\textbf{x}_{[i_1]}$ uses ``[MASK]'' but $\textbf{x}_{[i_2]}$ uses $x_{i_1}^*$. More general, the only one difference between $\textbf{x}_{[i_{j-1}]}$ and $\textbf{x}_{[i_j]}$ is that in the $i_{j-1}$-th position, $\textbf{x}_{[i_{j-1}]}$ uses ``[MASK]'' but $\textbf{x}_{[i_j]}$ uses $x_{i_{j-1}}^*$ for any $2\leq j\leq s$.

Taking Fig. 1 for explanation, the cover text is ``\emph{Midshire is a wonderful little city .}'' and there are two masked positions. For the first masked position (from left to right), the temporary text to be fed into the masked LM is ``\emph{Midshire is a [MASK] little [MASK] .}''. After the first masked position was replaced with the stego word ``\emph{nice}'', the temporary text to be fed into the masked LM for the second masked position is ``\emph{Midshire is a nice little [MASK] .}''. It is seen that the temporary text for the present masked position depends on the generation of the previous temporary text. We deem it an autoregressive method. 

\subsection{Consistency Coding}
A most important problem is how to build the mapping relationship between words and secret bits. To deal with this problem, the authors in \cite{HonaiUeoka:2021} use a threshold $0\leq t_p\leq 1$ to collect a list of candidate words whose prediction probabilities are higher than $t_p$, and then map each of the candidate words to a fixed-length binary stream. In other words, the collected candidate words have the same probability of being selected to fill in the present masked position, which does not take into account the probability distribution of the candidate words and may not fit into the context well. 

As mentioned previously, it is quite desirable to select such a word that it not only matches the secret bits to be embedded but also has a prediction probability obtained from the masked LM as high as possible. Because the secret bits are evenly distributed, the probability of selecting a specific word as the output actually depends on the number of secret bits carried by the word. In other words, for any two words mapped to a binary stream with the same length, they often have the same probability of matching the secret stream to be embedded. For example, in \cite{HonaiUeoka:2021}, if all candidate words are mapped to a binary stream with a length of $l > 0$, the probability of choosing any one among $2^l$ candidate words as the present output is $1/2^l$. It indicates that once the mapping relationship is finished, we are no longer free to choose the word. Therefore, it is necessary that the construction of mapping relationship itself has taken into account the prediction probability distribution so that for those words with a high prediction probability, the probability of choosing any one of them is high as well.

In summary, we expect to find such an information encoding strategy that the probability of choosing a word as the present output is proportional to its prediction probability obtained by the masked LM. We treat such information encoding strategy as \emph{consistency coding}. Compared with the  method introduced in \cite{HonaiUeoka:2021}, consistency coding has two significant advantages: 1) the number of candidate words can be any integer, rather than a power of two which is required in \cite{HonaiUeoka:2021}, and; 2) it bases on the statistical probability distribution of each word in the vocabulary, which takes into account the frequency of words and makes coding more conducive to the regularization of the text. Such a mechanism imitates the priority of native speakers in choosing words, and improves the security accordingly.

\begin{table*}[!t]
\caption{Example comparison between different linguistic steganographic methods due to different parameters.}
\centering
\begin{tabularx}{\linewidth}{c|c|X|c|c}
\hline\hline
$f$ & Type & Text & PPL & Payload\\
\hline
\multirow{8}{*}{2} & \multirow{2}{*}{Cover} & \emph{Gerson and Walter contend there are multiple  \textbf{paths} to solving a  \textbf{mathematical} problem and  \textbf{students} should be encouraged to  \textbf{chart} their own way by  \textbf{exploring} problems  \textbf{drawn} from the real  \textbf{world} .} & \multirow{2}{*}{-} & \multirow{2}{*}{-} \\
\cline{2-5} 
 & \multirow{3}{*}{\cite{HonaiUeoka:2021}}  & \emph{Gerson and Walter contend there are multiple \textbf{ways (2,1)}  to solving a \textbf{particular (8,001)} problem and \textbf{students (4,10)} should be encouraged to \textbf{go (4,00)} their own way by \textbf{exploring (0,-)} problems \textbf{different (8,001)} from the real \textbf{one (4,10)} .} & \multirow{3}{*}{63.5261} & \multirow{3}{*}{0.42} \\
\cline{2-5} & \multirow{3}{*}{Proposed} & \emph{Gerson and Walter contend there are multiple \textbf{approaches (3,1)} to solving a \textbf{given (7,001)} problem and \textbf{users (7,100)} should be encouraged to \textbf{find (4,000)} their own way by \textbf{solving (3,1)} problems \textbf{different (4,1)} from the real \textbf{one (5,000)} .} & \multirow{3}{*}{\textbf{56.4565}} & \multirow{3}{*}{\textbf{0.48}} \\
\hline
\multirow{8}{*}{3} & \multirow{2}{*}{Cover} & \emph{An unlucky accident befell my \textbf{servant} , Stevens , in falling from the \textbf{coach} and being dragged by the foot \textbf{upon} the pavement . Edition : current ; Page : [186] He was in great \textbf{danger} , but happily was not essentially \textbf{hurt} .} & \multirow{2}{*}{-} & \multirow{2}{*}{-} \\
\cline{2-5} 
 & \multirow{3}{*}{\cite{HonaiUeoka:2021}}  & \emph{An unlucky accident befell my \textbf{friend (4,00)} , Stevens , in falling from the \textbf{roof (8,101)} and being dragged by the foot \textbf{onto (2,1)} the pavement . Edition : current ; Page : [186] He was in great \textbf{shape (4,01)} , but happily was not essentially \textbf{happy (4,11)} .} & \multirow{3}{*}{105.5982} & \multirow{3}{*}{0.27} \\
\cline{2-5} & \multirow{3}{*}{Proposed} & \emph{An unlucky accident befell my \textbf{nephew (7,0010)} , Stevens , in falling from the \textbf{window (9,1101)} and being dragged by the foot \textbf{across (3,11)} the pavement . Edition : current ; Page : [186] He was in great \textbf{debt (8,000)} , but happily was not essentially \textbf{rich (7,1000)} .} & \multirow{3}{*}{\textbf{96.9807}} & \multirow{3}{*}{\textbf{0.46}} \\
\hline
\multirow{12}{*}{4} & \multirow{4}{*}{Cover} & \emph{Accountability is a major \textbf{key} to BILSTEIN’s success . We hold our team \textbf{members} , line employees , frontline leaders , \textbf{managers} and company leaders responsible for \textbf{thinking} above the line in addressing both success and \textbf{failure} , taking direct responsibility for \textbf{situations} through personal steps to solve and \textbf{overcome} issues and not becoming a victim for \textbf{individual} or collective results .} & \multirow{4}{*}{-} & \multirow{4}{*}{-} \\
\cline{2-5} 
 & \multirow{4}{*}{\cite{HonaiUeoka:2021}}  & \emph{Accountability is a major \textbf{key (4,01)} to BILSTEIN’s success . We hold our team \textbf{leaders (2,1)} , line employees , frontline leaders , \textbf{managers (4,00)} and company leaders responsible for \textbf{working (8,010)} above the line in addressing both success and \textbf{loss (4,01)} , taking direct responsibility for \textbf{situations (0,-)} through personal steps to solve and \textbf{address (4,10)} issues and not becoming a victim for \textbf{individual (2,0)} or collective results .} & \multirow{4}{*}{64.4565} & \multirow{4}{*}{0.23} \\
\cline{2-5} & \multirow{4}{*}{Proposed} & \emph{Accountability is a major \textbf{key (5,011)} to BILSTEIN’s success . We hold our team \textbf{managers (3,00)} , line employees , frontline leaders , \textbf{executives (5,010)} and company leaders responsible for \textbf{staying (11,01)} above the line in addressing both success and \textbf{failure (6,1)} , taking direct responsibility for \textbf{moving (6,0010)} through personal steps to solve and \textbf{resolve (5,0)} issues and not becoming a victim for \textbf{mistakes (3,01)} or collective results .} & \multirow{4}{*}{\textbf{57.0832}} & \multirow{4}{*}{\textbf{0.32}} \\
\hline\hline
\end{tabularx}
\label{table:examples}
\end{table*}

It is free for us to design the consistency coding. However, many classical entropy encoding methods in information theory can be used to realize consistency coding such as Huffman coding, arithmetic coding and Shannon-Fano coding \cite{information:theory}. For simplicity, we use Huffman coding to provide the off-the-shelf solution. Formally, for compactness, for each of the specific masked positions, let $\{p_1, p_2, ..., p_z\}$ be the prediction probabilities for the words in the vocabulary $V = \{v_1, v_2, ..., v_z\}$, where $p_1\geq p_2\geq ...\geq p_z$ and $\sum_{i=1}^{z}p_i = 1$. We first collect all the words whose prediction probabilities are higher than a pre-determined threshold $t_p$. The collected words are then regarded as the candidate words to be encoded. Then, by normalizing the prediction probabilities of the candidate words, we further apply Huffman coding to map each candidate word to a stream. For example, let $V' = \{v_1, v_2, ..., v_w\} \subset V,~(w \leq z)$, include the candidate words, the normalization operation is:
\begin{equation}
p_i \leftarrow \frac{p_i}{\sum_{j=1}^{w}p_j}, \forall 1\leq i\leq w,
\end{equation}
which enables us to thereafter directly apply Huffman coding. 

\emph{Remark 1:} When the entire secret message cannot be carried by a single cover text, multiple cover texts can be used. When the entire secret message can be carried by a single cover text, some masked positions in the cover text will be embedded with secret bits while the other masked positions (if any) are filled with the original words without data embedding. 

\emph{Remark 2:} It is known that Huffman coding does not result in the unique mapping. For example, in Fig. 1, the two words ``city'' and ``town'' are mapped to ``0'' and ``1'', respectively. It is also possible that  ``city'' is mapped to ``1'' and the other is mapped to ``0''. As long as the data hider and the data receiver use the same secret key to control the Huffman coding process, they can obtain the same mapping result. 

\section{Experimental Results and Analysis}
In the experiments, we use the large-scale CC-100 dataset \cite{CCNet:paper} consisting of high-quality data obtained through web crawling and containing monolingual data in more than 100 languages. A total of 10,000 texts are randomly selected from the English part of the CC-100 dataset, which guarantees the randomness and universality, therefore better proves the reliability of the proposed method. One thing to note is that the lengths of the cover texts are required to be larger than 20 so that a sufficient payload can be embedded. The cover texts are embedded with a randomly generated binary stream to generate the stego texts. As mentioned in Section III, we use BERT as the masked LM. For comparison, following \cite{HonaiUeoka:2021}, we use the Google's $\text{BERT}_\text{Base, Cased}$ model and Hugging Face's transformers package \cite{Transformer:paper} with the default settings. In other words, our setup closely matches \cite{HonaiUeoka:2021} for fair comparison. In addition, we use the commonly used indicator perplexity (PPL) \cite{Yang:paper} to evaluate the text quality and measure the security by steganalysis. In general, a lower PPL indicates that the quality of the text is better. 

\begin{figure}[!t]
\centering
\includegraphics[width=\linewidth]{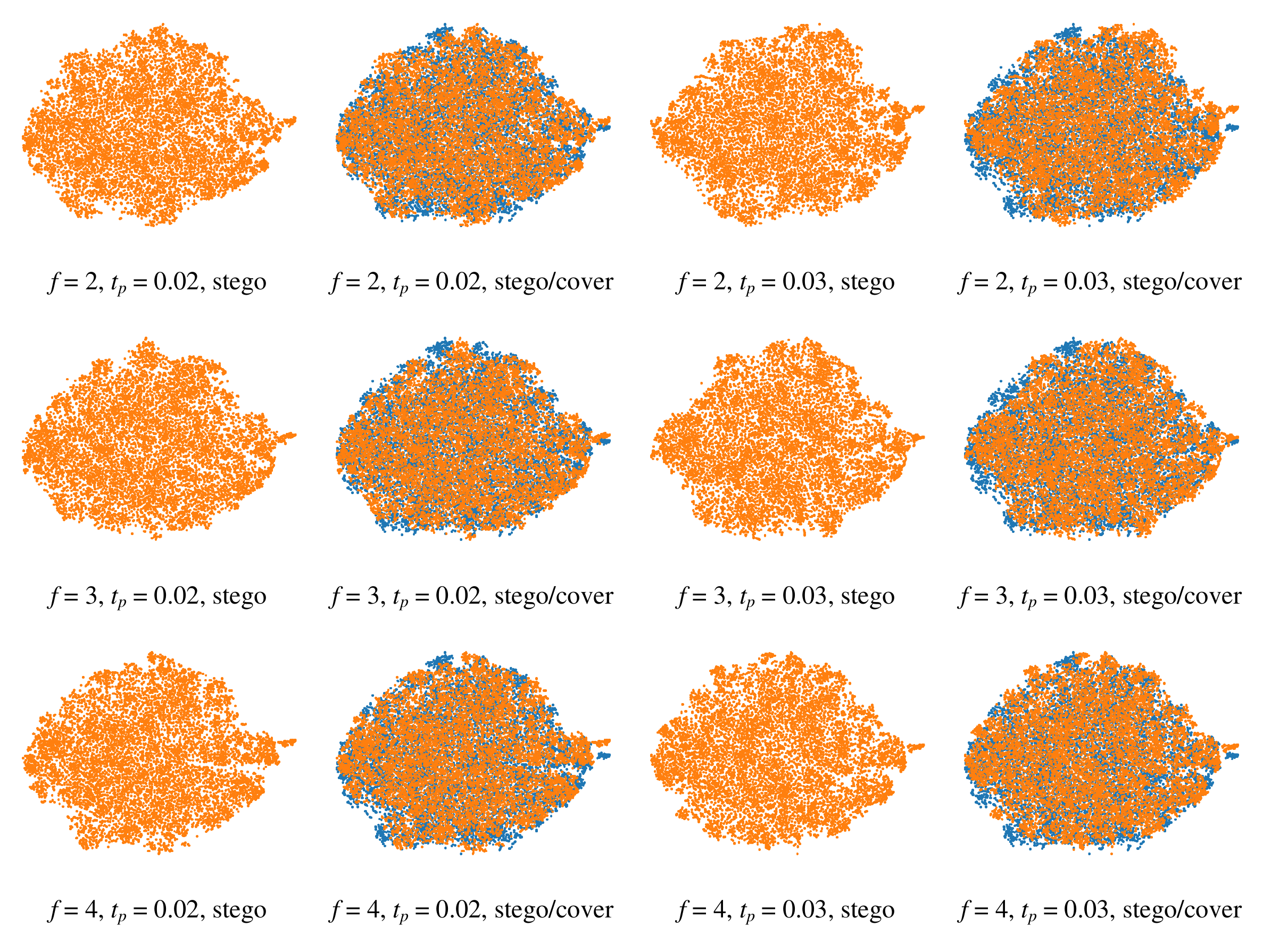}
\caption{Visualization for stego texts and natural ones by applying t-SNE \cite{tsne:paper}. The blue dots represent natural texts and the others are stego texts. }
\end{figure}

\subsection{Qualitative Results}
We first show some examples to evaluate the quality of the stego texts generated by the proposed method. Table \ref{table:examples} shows the experimental examples, in which we compare the proposed method with \cite{HonaiUeoka:2021} due to the close correlation mentioned in Section II. The threshold $t_p$ defined in Section II and III is empirically set to 0.02 by default. The embedding parameter $f$ controls the number of the masked positions. The higher $f$ is, the lower the number of the masked positions is. The payload is measured by bits per word (bpw). Since the masking strategy always skips punctuation, the punctuation marks never carry data. For simplicity, the punctuation marks are excluded when to determine the total number of words in a text, e.g., in Table \ref{table:examples}, when $f = 2$, the payload of the stego text generated by \cite{HonaiUeoka:2021} is 13/31 = 0.42, rather than 13/32 = 0.41. In Table \ref{table:examples}, ``$W(a,b)$'' indicates that the word ``$W$'' is mapped to a binary stream $b$ and the number of the candidate words for the present masked position is $a$ during data embedding. For example, ``ways(2,1)'' means that the word ``ways'' carries only one secret bit ``1'' and there are 2 candidate words for the present position. ``exploring (0,-)'' means that there has no candidate words for choice and the original word ``exploring'' is used for the present position. 

It can be inferred from Table \ref{table:examples} that compared with \cite{HonaiUeoka:2021}, the proposed method not only improves the semantic quality, but also increases the payload for all the three cases, which have verified the superiority of the proposed method. This can be explained from two aspects. On the one hand, compared with the non-autoregressive prediction strategy in \cite{HonaiUeoka:2021}, the proposed autoregressive prediction strategy enhances the semantic connection between contexts so that those words better fitting into contexts can be collected for data embedding. On the other hand, compared with block coding \cite{HonaiUeoka:2021}, the proposed consistency coding strategy is based on the prediction probability distribution of each word in the vocabulary, which enlarges the candidate set and makes the choice of stego words more conducive to maintaining the text quality. 

\begin{table}[!t]
\renewcommand{\arraystretch}{1.2}
\caption{Detection accuracy for detecting different steganographic methods with different payloads.}
\centering
\begin{tabular}{c|c|c|c}
\hline\hline
Method & Parameters & Mean Payload & Accuracy\\
\hline
Ref. \cite{HonaiUeoka:2021} & $f = 3, t_p = 0.02$ & 0.2479 & 0.5570\\
Ref. \cite{Ziegler:paper} & $\tau = 0.7$ & 2.2934 & 0.8967\\
Ref. \cite{Yang:paper} & $L = 2^2$ & 1.8072 & 0.9580\\
Ref. \cite{Fang:paper} & $B = 2^2$ & 2.0000 & 0.9615\\
Proposed & $f = 3, t_p = 0.02$ & 0.2506 & \textbf{0.5498}\\
\hline\hline
\end{tabular}
\end{table}

\begin{table}[!t]
\renewcommand{\arraystretch}{1.2}
\centering
\caption{Detection accuracy for detecting different steganographic methods with different masking intervals. ($t_p = 0.02$)}
\begin{tabular}{c|c|c|c|c}
\hline\hline
\multicolumn{1}{c|}{\multirow{2}{*}{$f$}} &
\multicolumn{2}{c|}{Ref. \cite{HonaiUeoka:2021}} & \multicolumn{2}{c}{Proposed} \\
\cline{2-5}
& Mean Payload & Accuracy  &  Mean Payload & Accuracy \\
\hline
$2$  & 0.3818 & 0.6107 & \textbf{0.3906} & \textbf{0.6045}\\
$3$  & 0.2479 & 0.5570 & \textbf{0.2506} & \textbf{0.5498}\\
$4$ & 0.1903 & 0.5403 & \textbf{0.1937} & \textbf{0.5337}\\
\hline\hline
\end{tabular}
\end{table}

\begin{table}[!t]
\renewcommand{\arraystretch}{1.2}
\centering
\caption{Detection accuracy for detecting different steganographic methods with different masking intervals. ($t_p = 0.03$)}
\begin{tabular}{c|c|c|c|c}
\hline\hline
\multicolumn{1}{c|}{\multirow{2}{*}{$f$}} &
\multicolumn{2}{c|}{Ref. \cite{HonaiUeoka:2021}} & \multicolumn{2}{c}{Proposed} \\
\cline{2-5}
& Mean Payload & Accuracy  &  Mean Payload & Accuracy \\
\hline
$2$  & 0.2916 & 0.6050 & \textbf{0.3145} & \textbf{0.5968}\\
$3$  & 0.1919 & 0.5532 & \textbf{0.2057} & \textbf{0.5465}\\
$4$ & 0.1474 & 0.5417 & \textbf{0.1578} & \textbf{0.5407}\\
\hline\hline
\end{tabular}
\end{table}

\begin{table*}[!t]
\renewcommand{\arraystretch}{1}
\centering
\caption{Performance comparison between different steganographic strategies. The Payload and PPL are using mean values.}
\begin{tabular}{c|c|c|c|c|c|c|c|c|c|c|c|c|c}
\hline\hline
\multicolumn{1}{c|}{\multirow{2}{*}{$f$}} &
\multicolumn{1}{c|}{\multirow{2}{*}{$t_p$}} &
\multicolumn{3}{c|}{Ref. \cite{HonaiUeoka:2021}} &
\multicolumn{3}{c|}{Autoregressive only} &
\multicolumn{3}{c|}{Consistency coding only} &
\multicolumn{3}{c}{Proposed}
\\
\cline{3-14}
& & Payload & PPL & Accuracy  &  Payload & PPL & Accuracy &  Payload & PPL & Accuracy &  Payload & PPL & Accuracy\\
\hline
$2$  & \multirow{3}{*}{$0.02$} & 0.3818	& 86.0537 & 0.6107 & 0.3818 & 84.8311 & 0.6048 & \textbf{0.3916} & 83.1867 & 0.6123 & 0.3906 & \textbf{82.0609} & \textbf{0.6045}\\
$3$  &  & 0.2479 & 81.7708 & 0.5570 & 0.2473 & 81.5283 & \textbf{0.5487} & \textbf{0.2532} & 79.7653 & 0.5612 & 0.2506 & \textbf{79.4073} & 0.5498\\
$4$ &  & 0.1903 & 80.6252 & 0.5403 & 0.1897 & 80.3875 & 0.5398 & \textbf{0.1938} & 78.9204 & 0.5423 & 0.1937 & \textbf{78.9092} & \textbf{0.5337}\\
\hline
$2$  & \multirow{3}{*}{$0.03$} & 0.2916 & 82.7407 & 0.6050 & 0.2934 & 81.7111 & 0.6028 & 0.3134 & 81.3915 & 0.6045 & \textbf{0.3145} & \textbf{80.4290} & \textbf{0.5968}\\
$3$  &  & 0.1919 & 79.4869 & 0.5532 & 0.1923 & 79.3132 & 0.5512 & \textbf{0.2058} & 78.5438 & 0.5520 & 0.2057 & \textbf{78.3508} & \textbf{0.5465}\\
$4$ &  & 0.1474 & 78.8377 & 0.5417 & 0.1472 & 78.7971 & 0.5410 & \textbf{0.1578} & \textbf{78.2237} & 0.5433 & \textbf{0.1578} & 78.2437 & \textbf{0.5407}\\
\hline\hline
\end{tabular}
\end{table*}

\subsection{Quantitative Results}
To evaluate the performance against steganalysis, following \cite{HonaiUeoka:2021}, for each experiment, we first split the above-mentioned 10,000 natural texts and their stego versions to three disjoint subsets, i.e., training set (60\%), validation set (10\%), and testing set (30\%), and then finetune the $\text{BERT}_\text{base-cased}$ model with the training set and the validation set. The finetuned model is used to detect whether a text to be tested is stego or not. The accuracy is evaluated on the testing set with the model of the highest validation accuracy. We compare the proposed work with four state-of-the-art methods, i.e., \cite{HonaiUeoka:2021}, \cite{Ziegler:paper}, \cite{Yang:paper} and \cite{Fang:paper}. Table II shows the experimental results. In Table II, the parametric settings are described as follows. Ref. \cite{HonaiUeoka:2021} and our work use the same $f = 3$ and $t_p = 0.02$ for fair comparison. The explanation for $f$ and $t_p$ has been described in Subsection IV-A. $\tau$ is a system parameter introduced in \cite{Ziegler:paper} and we use the recommended value in \cite{Ziegler:paper}, i.e., $\tau = 0.7$. The authors in \cite{Yang:paper} propose two methods called fixed-length coding and variable-length coding for information encoding. We use the variable-length coding strategy for simulation since it has the better performance. For \cite{Yang:paper} and \cite{Fang:paper}, the truncation length $L$ and the block size $B$ are set to $2^2$ for fair comparison as well. As shown in Table II, different methods have different performance. In terms of payload, the methods in \cite{Ziegler:paper}, \cite{Yang:paper}, \cite{Fang:paper} outperform \cite{HonaiUeoka:2021} and the proposed method. The reason is that the three methods are actually generation based while \cite{HonaiUeoka:2021} and the proposed method are actually modification based. Generation based methods often provide a high payload since each word can be used to carry secret information. However, in Table II, the detection accuracies of the three generation based methods are high, meaning that the security is not satisfied. By comparing with \cite{HonaiUeoka:2021}, we can find that the proposed work not only achieves a higher payload, but also has a lower detection accuracy, indicating that the proposed work achieves a better trade-off between payload and security.

To further demonstrate the superiority of the proposed work, we visualize the statistical distribution of the stego texts and the natural (cover) texts by applying t-SNE \cite{tsne:paper}. It is observed from Fig. 2 that when the payload (i.e., BPW) decreases, more points of the two colors overlap, implying that the stego texts are more approximate to the natural ones and therefore indeed have higher security. For the different parameters given in Fig. 2, we also show the corresponding mean payload and detection accuracy for \cite{HonaiUeoka:2021} and the proposed work. As shown in Table III and Table IV, the proposed work outperforms \cite{HonaiUeoka:2021} for all cases, which has verified the superiority of the proposed work. We also determine the mean PPL of stego texts due to different parameters. The experimental results are plotted in Fig. 3. It can be inferred that the proposed work also generate stego texts with better quality than \cite{HonaiUeoka:2021}, indicating that it is more difficult for a person to distinguish between stego texts generated by the proposed work and natural texts compared with \cite{HonaiUeoka:2021}, which reveals that the proposed work has high-level concealment. 

\begin{figure}[!t]
\centering
\includegraphics[width=\linewidth]{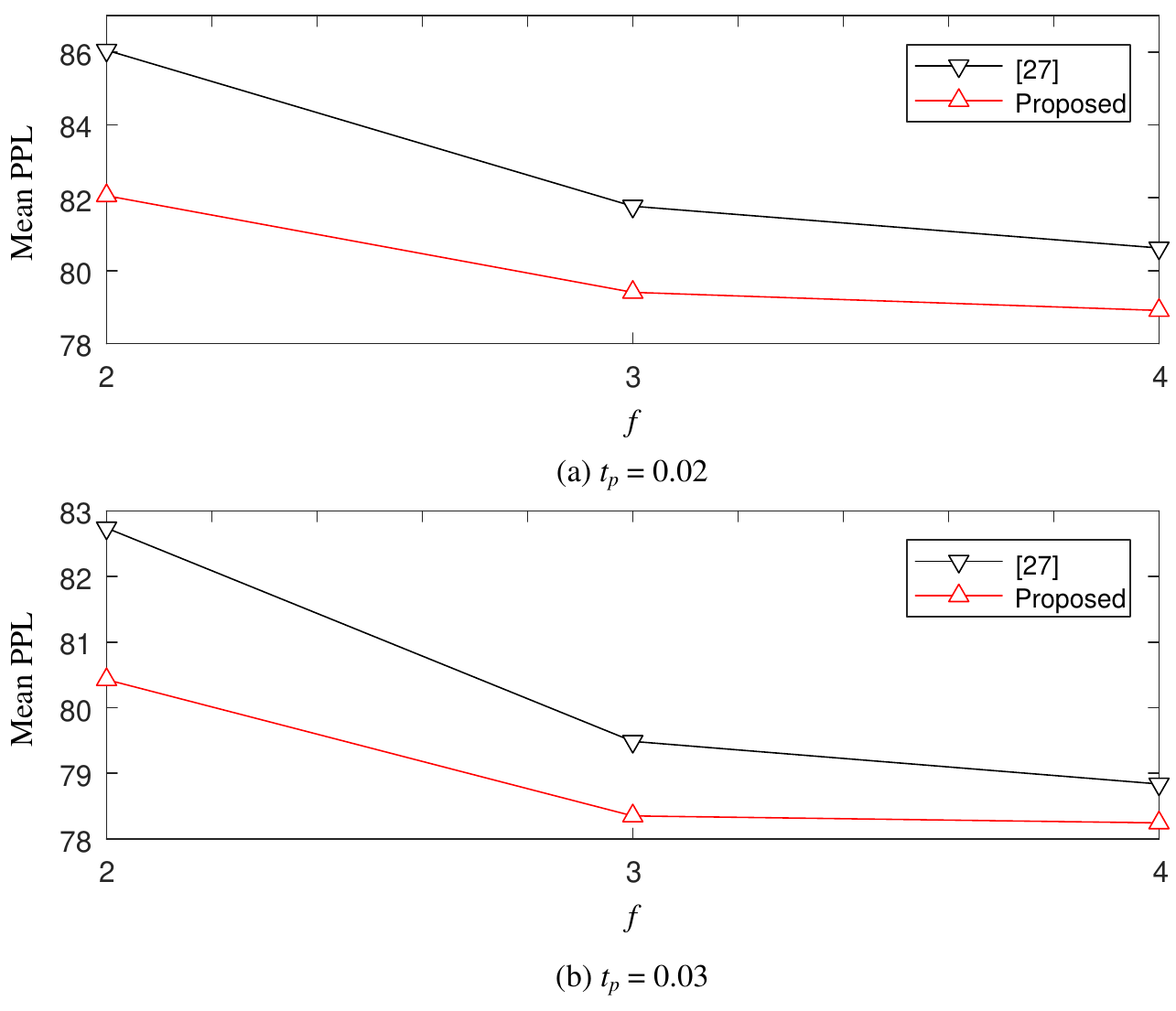}
\caption{The mean PPLs for the stego texts generated by different methods.}
\end{figure}

\subsection{Ablation Study}
The proposed work achieves superior performance by using an autoregressive strategy and a consistency coding technique. In order to show that both the autoregressive strategy and the consistency coding technique have the ability to improve the security or the payload, we further conduct ablation study for evaluation. The method in \cite{HonaiUeoka:2021} use the non-autoregressive strategy and block coding technique for steganography. By substituting non-autoregressive with autoregressive, we can develop a new steganographic method named as ``Autoregressive only''. Similarly, by substituting block coding with consistency coding, a new steganographic method named as ``Consistency coding only'' can be simulated. Table V shows the experimental results. In terms of payload, ``Consistency coding only'' and the proposed work are close to each other, but both higher than \cite{HonaiUeoka:2021} and ``Autoregressive only''. It indicates that the proposed work can benefit from the consistency coding technique so as to increase the payload. In terms of steganalysis accuracy, ``Autoregressive only'' and the proposed work have the similar performance, but both outperform \cite{HonaiUeoka:2021} and ``Consistency coding only''. It means that the autoregressive strategy is more secure than the non-autoregressive strategy. By combining the autoregressive strategy and the consistency coding technique, the PPL value of the proposed work is the lowest in most cases. In summary, it can be concluded that the proposed work has achieved the best performance compared with related works. 

\section{Conclusion and Discussion}
In this paper, we propose a novel autoregressive linguistic steganographic algorithm based on the BERT model equipped with consistency coding. By applying the autoregressive word-prediction strategy, the generated stego words better fit into the context, which improves the readability and the authenticity of the generated stego text. Meanwhile, the consistency coding technique enlarges the number of candidate words for LS and makes use of the statistical probability distribution of candidate words, which not only enhances the payload but also makes the stego text seemingly more natural. Experiments show that the proposed work has achieved the best performance compared with related works, verifying the superiority and applicability.  

From a practical perspective, with the widespread use of text over social networks, LS will become more and more popular. By conveying a stego text through a social network platform, the steganographic behavior can be easily concealed by the huge number of ordinary social activities. More importantly, once the receiver observes the stego text, he can keep silent and extract the secrets without taking any suspicious interaction, which conceals the real identify of the receiver. Though there are increasing LS methods that are reported in the literature in recent years, how to achieve a good trade-off between payload, text quality and security is still a challenging problem. On the one hand, though modification based LS well maintains the semantic quality of the text, the embeddable payload is low. On the other hand, though generation based LS allows us to embed a high payload, we need to pay the high price of uncontrollable semantic information and security. In this paper, by exploiting a language model for word prediction and making use of the probability distribution for information encoding, the proposed work improves the fluency of the stego text while guaranteeing security compared with modification based methods, and improves the security compared with generation based methods. The payload is increased as well to a certain extent, which brings modification based LS closer to generation based LS. In the future, we will explore efficient strategies to increase the embeddable payload to build a bridge between modification based LS and generation based LS. 

\section*{Acknowledgement}
This work was supported by the National Natural Science Foundation of China under Grant number 61902235, and the Shanghai Chenguang Program under Grant number 19CG46.

%\ifCLASSOPTIONcaptionsoff
%  \newpage
%\fi

\end{document}